\documentstyle[sprocl,epsfig,epsf]{article}

\def\ra{\rightarrow}
\def\be{\begin{equation}}
\def\ee{\end{equation}}
\def\bea{\begin{eqnarray}}
\def\eea{\end{eqnarray}}

\def\prw{\widetilde{\Pi}_{\rho\omega}}
\def\rhoom{$\rho^0$-$\omega\;$}
\def\pieta{$\pi^0$-$\eta,\eta'\;$}
\def\Brwpi{$B^{\mp} \ra \rho^{\mp} \rho^0(\omega) 
\ra \rho^{\mp} \pi^+ \pi^-\;$}

\def\Bpipi{$B \ra \pi\pi\;$}
\def\etwopi{$e^+e^-\ra\pi^+\pi^-\;$}

\def\lapp{\hbox{$ {     \lower.40ex\hbox{$<$}
                   \atop \raise.20ex\hbox{$\sim$}
                   }     $}  }
\def\rapp{\hbox{$ {     \lower.40ex\hbox{$>$}
                   \atop \raise.20ex\hbox{$\sim$}
                   }     $}  }

\newcommand{\CP}{\not \! \! {\rm CP}}

\begin{document}
{\vbox{
                                       \hfill UK/TP 98-14  \\
                                       \null\hfill hep-ph/9809479\\
                                       \null\hfill September 1998}}

\title{ISOSPIN VIOLATION IN HADRONIC B-DECAYS
\footnote{To appear in the proceedings of the 
{\it Workshop on CP Violation}, Adelaide, Australia,
July 3-8 1998.}}

\author{S. GARDNER}

\address{Department of Physics and Astronomy, University of Kentucky, \\
Lexington, KY 40506-0055 USA\\E-mail: svg@ratina.pa.uky.edu} 


\maketitle\abstracts{
I present two disparate examples of isospin violation in hadronic
B-decays. In the first, the presence of 
\rhoom mixing in the decay \Brwpi permits 
the extraction of sgn($\sin\alpha$), where
$\alpha$ is the usual angle of the unitarity triangle, 
with minimal hadronic uncertainty.
In the second, the presence of \pieta mixing can obscure
the extraction of $\sin 2\alpha$ from an isospin analysis in
\Bpipi decays.}

%
%

\vspace{-1cm}
\section{Introduction}

In the standard model, 
CP violation is characterized by a single phase in the 
Cabibbo-Kobayashi-Maskawa (CKM) matrix, rendering its elements complex. 
The CKM matrix of the standard model is unitary, so that 
determining whether or not this is empirically so is a central test of the
standard model's veracity~\cite{flrev96}. Thus, determining whether
the angles of the unitarity triangle, 
$\alpha$, $\beta$, and $\gamma$, empirically sum to $\pi$
and whether its angles are compatible with the measured lengths of
its sides lie at the heart of these tests of the standard model. 

CP-violating effects in hadronic B-decays 
will play a crucial role in the determination
of $\alpha$, $\beta$, and $\gamma$, and many 
clever methods have been devised 
to evade the uncertainties the
strong interaction would weigh on their extraction~\cite{flrev96}. 
Irrespective of the efficacy of these methods, 
discrete ambiguities in the angles 
remain, for experiments in 
the neutral B sector which would measure 
an unitarity triangle angle, $\phi$, 
determine $\sin 2 \phi$ and not
$\phi$ itself~\cite{GQ97}. 
Nevertheless, 
removing discrete ambiguities is important,
for standard model unitarity requires merely
$\alpha + \beta + \gamma= \pi$, mod $2\pi$. Determining
the precise equality
yields another
standard model test, for 
consistency with the measured value of $\epsilon$ and 
the computed $B_K$ parameter suggest that it ought
be $\pi$~\cite{NQ90}. 

The isospin-violating effects to be discussed here impact 
the extraction of $\alpha$, where 
$\alpha\equiv 
{\rm arg} [-V_{td} V_{tb}^\ast/(V_{ud}V_{ub}^\ast)]$
and $V_{ij}$ is an element of the CKM matrix~\cite{pdg96}.
The first exploits isospin violation
to extract sgn($\sin \alpha$) from the rate asymmetry in \Brwpi, where
$\rho^0(\omega)$ denotes 
the \rhoom interference region~\cite{ET96,got97}, with
minimal hadronic uncertainty, removing the mod($\pi$) ambiguity in
$\alpha$ consequent to a $\sin 2\alpha$ measurement. 
An asymmetry emerges only if 
both a weak and strong phase difference exists between two 
interfering amplitudes~\cite{bss79}. The 
strong phase difference is typically
small and uncertain, yet in the decay
\Brwpi,
the presence of the $\omega$ resonance not only enhances the
asymmetry to some 20\%~\cite{ET96,got97} 
but also permits the determination of the strong phase from
\etwopi data for $\pi^+\pi^-$ invariant masses 
in the vicinity of the $\omega$ resonance~\cite{got97}.

The second topic 
concerns the impact of isospin violation on the 
extraction of $\sin 2\alpha$ from an isospin analysis 
in \Bpipi decays~\cite{GL90}. Isospin is broken not only by 
electroweak effects but also by 
the $u$ and $d$ quark mass difference. The latter drives
$\pi^0-\eta,\eta'$ mixing~\cite{leut96}, which, in turn, generates an
amplitude in \Bpipi not included in the isospin analysis. Thus, although 
the effect of electroweak penguins is estimated
to be small~\cite{deshe95,gronau95,fleischer96}, 
when all the effects of isospin violation are included, 
the error in the extracted value of $\sin 2\alpha$ can 
be significant~\cite{pieta98}. 

%
%

\vspace{-3mm}
\section{\rhoom Mixing and CP Violation 
in $B^{\mp}\ra \rho^{\mp}\rho^0(\omega)\ra \rho^{\mp}\pi^+\pi^-$}

Here we consider the extraction of sgn($\sin \alpha$) from 
\Brwpi, as proposed by
Enomoto and Tanabashi~\cite{ET96}. 
In this channel, the rate asymmetry, which is CP-violating, 
is also isospin forbidden. 
If isospin were 
a perfect symmetry, then the Bose symmetry
of the $J=0$ $\rho^\pm \rho^0$ final state would force it to have
isospin $I=2$. The strong penguin
amplitude is also purely $\Delta I=1/2$, so that no CP violation is 
possible in this limit. 
If isospin violating effects are 
included, however, two effects occur. The penguin operators then
possess both $\Delta I=1/2$ and $\Delta I=3/2$ character; the latter are
generated by electroweak penguin operators and by the 
isospin-violating effects which distinguish the $\rho^\pm$ from the
$\rho^0$. Yet isospin violation also generates \rhoom mixing,
so that a $I=1$ final state is also possible. In our detailed
numerical estimates~\cite{got97}, we find
that the strong phase in the \rhoom interference region is 
driven by \rhoom mixing. The sign and magnitude of \rhoom mixing 
is fixed by \etwopi data~\cite{miller90,pionff97}, so that we
are able to interpret direct CP violation in this channel
to extract sgn($\sin\alpha$). 

Resonances can play a strategic role 
in direct CP violation, for their mass and
width can be used 
to constrain the strong phase and their interference can significantly
enhance the CP-violating asymmetry~\cite{atwood95,eilam95}. 
To see how these effects are realized here, consider the amplitude 
$A$ for $B^- \rightarrow \rho^- \pi^+ \pi^-$ decay:
$A = \langle \pi^+\pi^- \rho^- | {\cal H}^{\rm T} | B^- \rangle
+ \langle \pi^+\pi^- \rho^- | {\cal H}^{\rm P} | B^- \rangle$, 
where $A$ is given by the sum of the amplitudes corresponding to the tree
and penguin diagrams, respectively. Defining the strong phase
$\delta$, the weak phase $\phi$, and the magnitude $r$ via 
$A = \langle \pi^+\pi^- \rho^- | {\cal H}^{\rm T} | B^- \rangle [
1 + re^{i\delta}\; e^{i\phi} ]$
the CP-violating 
asymmetry $A_{\;\CP}$ is 
\be
A_{\;\CP} \equiv 
{ | A |^2 - |{\overline A}|^2 
\over | A |^2 + |{\overline A}|^2 }
= {-2r \sin \delta \sin \phi 
\over 1 + 2r\cos \delta \cos \phi + r^2 } \;,
\label{asym}
\ee
where $\phi$ is $- \alpha$~\cite{pdg96}.
To express $\delta$ in terms of the resonance
parameters, 
let $t_{\rm V}$ be the tree amplitude 
and $p_{\rm V}$ be the penguin amplitude to produce a vector meson ${\rm V}$. 
Thus, the tree and penguin 
amplitudes 
become
\bea
\langle \pi^+\pi^- \rho^- | {\cal H}^{\rm T} | B^- \rangle 
= {g_{\rho} \over s_\rho s_\omega} \prw t_{\omega}
  + { g_{\rho} \over s_\rho } t_\rho  \;,\\
\langle \pi^+\pi^- \rho^- | {\cal H}^{\rm P} | B^- \rangle 
= {g_{\rho} \over s_\rho s_\omega} \prw p_{\omega}
  + { g_{\rho} \over s_\rho } p_\rho \;, \label{fun}
\eea
where we have introduced
$\prw$, the effective 
$\rho^0$-$\omega$ mixing matrix
element~\cite{pionff97}; 
 $g_\rho$, the $\rho^0 \rightarrow \pi^+\pi^-$ coupling; $1/s_{\rm V}$,
the vector meson propagator, with 
$s_{\rm V}=s - m_{\rm V}^2 + i m_{\rm V} \Gamma_{\rm V}$;
and $s$, the square of the invariant mass of the $\pi^+ \pi^-$ pair. Using 
\be
re^{i\delta}\,e^{i\phi}= 
\frac{\langle \pi^+\pi^- \rho^- | {\cal H}^{\rm P} | B^- \rangle}
{\langle \pi^+\pi^- \rho^- | {\cal H}^{\rm T} | B^- \rangle}
\ee
and~\cite{ET96}:
\be
{p_\omega \over t_\rho} \equiv r' e^{i(\delta_q + \phi)} \;, \quad
{t_\omega \over t_\rho} \equiv {\tilde \alpha} e^{i \delta_{\tilde{\alpha}}} \;, \quad
{p_\rho \over p_\omega} \equiv \beta e^{i \delta_\beta} \;,
\label{alphabeta}
\ee
one finds, to leading order in isospin violation, 
\be
re^{i\delta} = 
{r' e^{i\delta_q}\over s_\omega} \left\{
 \prw + \beta e^{i\delta_\beta}
\left( s_\omega 
- \prw \tilde{\alpha} e^{i\delta_{\tilde{\alpha}}} \right)
\right\} 
\;.
\label{thescoop}
\ee
A $J=0,\,I=1$ $\rho^\pm\rho^0$ 
final state is forbidden by Bose symmetry if isospin is perfect, 
so that
$\beta$ is non-zero only if electroweak penguin contributions and
isospin violation in the $\rho^\pm$ and $\rho^0$ hadronic form
factors are included. Numerically, 
$|\prw|/(m_\omega \Gamma_\omega) \gg \beta$~\cite{got97}.
Thus, as $s\ra m_\omega^2$, 
\be
re^{i\delta} \ra \frac{r' e^{i\delta_q} \prw }{i m_\omega \Gamma_\omega}\;,
\ee
so that $\delta\ra \delta_q - \pi/2$ and 
$r\ra r'\prw /m_\omega \Gamma_\omega$ in this limit.
The asymmetry 
is thus
determined by the resonance parameters
$\prw$, $m_\omega$, and $\Gamma_\omega$; the
weak phase $\phi$; 
and the
``short distance'' parameters $r'$ and $\delta_q$. 
The latter are calculable within the context of the 
operator product expansion if the factorization approximation is
applied, though a ratio of hadronic form factors
enters as well. Here that ratio is modified from unity only by 
isospin-violating effects which distinguish the $\rho^\pm$ from
the $\rho^0$; its deviation from unity is no greater than 1\%~\cite{got97}. 
The presence of \rhoom mixing 
implies that the asymmetry as $s\ra m_\omega^2$ depends on
$\cos\delta_q$, {\it not} $\sin\delta_q$, so that 
its magnitude is insensitive to the value of the small parameter
$\delta_q$. 
The asymmetry for various $N_c$ and $k^2$, where $k$ is the momentum
of the virtual boson, 
as a function of $s$ is shown in Fig.~1, 
where we have used the effective Hamiltonian of 
Deshpande and He~\cite{deshe95}, in QCD to 
next-to-leading logarithmic (NLL) order,
in the factorization approximation~\cite{note}.

\begin{figure}
\label{fig_asym}
\centerline{\epsfig{file=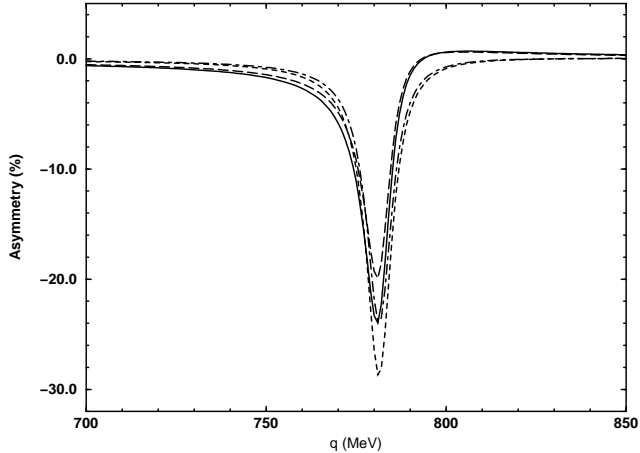,angle=-90,width=2.2in}}
\vspace{0.3in}
\caption{The CP-violating asymmetry, Eq.~[\protect{\ref{asym}}], in percent,
plotted versus the invariant mass $q$ of the $\pi^+\pi^-$ pair in MeV for
[$N_c$, $k^2/m_b^2$].
 The asymmetries with $\prw =-3500\;{\rm MeV}^2$ and 
$\tilde\varepsilon=-0.005$ are shown for [2, 0.5] (solid line), 
[3, 0.5] (long-dashed), [2, 0.3] (dashed), and [3, 0.3] (dot-dashed).}
\vspace{-5mm}
\end{figure}

To extract the sign of $\sin\alpha$, note that for
$s\approx m_\omega^2$ we have
\be
r \sin\delta
\approx \frac{\prw r'}{|s_\omega|^2} 
\left[ (s - m_\omega^2) \sin \delta_q - m_\omega \Gamma_\omega \cos \delta_q
\right] \;;
\label{rsind}
\ee
the sign of $\sin\delta$ at $s=m_\omega^2$ is that of 
$-\prw\cos\delta_q$. The sign and magnitude of $\prw$ is determined 
from a fit to the time-like pion form factor, $F_\pi(s)$, as measured 
in \etwopi, where~\cite{pionff97}
\be
F_\pi(s) = F_\rho(s) \left[ 1 + \frac{1}{3} \frac{\prw(s)}{s-m_\omega^2 
+ i m_\omega \Gamma_\omega}\right] \;.
\ee
Fitting \etwopi data we find 
$\prw = -3500\pm 300$ MeV$^2$~\cite{pionff97}. 
The error is statistical only; the value of $\prw$ 
is insensitive to the theoretical ambiguities in the $\rho$ 
parametrization, $F_\rho(s)$~\cite{pionff97}. Finally, then,
the determination of sgn($\sin\alpha$) relies on that
of sgn$(\cos\delta_q)$. The latter cannot be determined
without additional theoretical input; however, it is worth noting
that, in our computations in the factorization approximation,
$\cos\delta_q < 0$ for all $N_c$ and $k^2$~\cite{got97}. 
The situation, finally, is not dissimilar from that of the
determination of sgn($\sin 2\alpha$) in neutral B decays 
to CP eigenstates, for fixing the sign of $\sin 2\alpha$ requires
that of the parameter $B_B$~\cite{GKN97}. 
Nevertheless, the computation of 
$\cos\delta_q$ can be tested through the measurement of
the ``skew''  of the asymmetry 
in Eqs.~(\ref{asym}) and (\ref{rsind})
from a Breit-Wigner shape, noting Fig.~1. 
Indeed, the shape of the
asymmetry yields $\tan\delta_q$ and thus offers 
a sensitive test of factorization~\cite{got97}. 
This is possible as other effects which would skew the asymmetry 
are smaller still. 
For example, we have assumed $\prw$ to be both real and
$s$-independent, yet if we include these effects in our fits 
to \etwopi data, the phase and $s$-dependence of $\prw$
in the \rhoom interference region are statistically
consistent with zero~\cite{pionff97} and, moreover, do not mar our
interpretation of the skew~\cite{got97}. 

In summary, the rate asymmetry in \Brwpi is large and 
robust with respect to the known strong interaction 
uncertainties~\cite{guot98}. 
The presence of isospin violation in this decay permits the 
determination of 
sgn( $\sin\alpha$) once the sign of $\cos\delta_q$ is known, 
noting $|\cos\delta_q| \sim {\cal O}(1)$. 
The latter can be calculated, 
yet the shape of the asymmetry yields a direct test of the
suitability of our estimate.
$\tan\delta_q$ can
also be extracted from a comparison with the decay
$B^\pm \ra\rho^\pm\omega \ra\rho^\pm
\pi^+ \pi^0 \pi^-$; here the asymmetry goes as $\sin\delta_q$~\cite{got98}.
The sign of $\sin\alpha$ can be thus extracted with minimal hadronic 
uncertainty,
removing the mod($\pi$) ambiguity in $\alpha$ consequent to
a $\sin 2\alpha$ measurement. 

%
%

\vspace{-3mm}
\section{\pieta Mixing in $B\ra \pi\pi$ Decays}

  To review the isospin analysis
in $B\ra \pi\pi$ decays, due to Gronau and London~\cite{GL90}, 
let us consider 
the time-dependent asymmetry $A(t)$~\cite{pdg96}:
\be
A(t) = {( 1 - |r_{f_{CP}}|^2) \over ( 1 + | r_{f_{CP}}|^2)}
\cos(\Delta m\, t) 
- {2 ({\rm Im}\, r_{f_{CP}}) 
\over 
( 1 + | r_{f_{CP}}|^2)}
\sin (\Delta m\, t) \;, 
\ee
where $r_{f_{CP}} = ({V_{tb}^\ast V_{td} / V_{tb}V_{td}^\ast})
({{\overline A}_{f_{CP}}/ A_{f_{CP}}}) \equiv 
e^{-2i\phi_m} {{\overline A}_{f_{CP}} \over A_{f_{CP}}} $,
$A_{f_{CP}}\equiv A(B_d^0\ra f_{CP})$, and 
$\Delta m\equiv B_H - B_L$. 
Denoting the
amplitudes  $B^+ \ra \pi^+ \pi^0$, 
$B^0 \ra \pi^0 \pi^0$, and $B^0 \ra \pi^+ \pi^-$
by $A^{+0}$, $A^{00}$, and $A^{+-}$, respectively, 
and introducing $A_I$ to denote an
amplitude of final-state isospin $I$, we have~\cite{GL90}
\be 
{1\over 2} A^{+-} = A_2 - A_0 \quad;  A^{00} = 2A_2 + A_0 \quad; 
{1\over \sqrt{2}} A^{+0} =3 A_2 \;,
\label{triangle}
\ee
where analogous relations exist for 
$A^{-0}$, ${\overline A}^{00}$, and ${\overline A}^{+-}$
in terms of ${\overline A}_2$ and ${\overline A}_0$. If isospin
is perfect,
the Bose symmetry of the $J=0$ $\pi\pi$ state permits amplitudes 
of $I=0,2$, so that 
the amplitude $B^\pm \ra \pi^\pm \pi^0$ is purely $I=2$. Moreover,
the strong penguin contributions are 
of $\Delta I=1/2$ character, so that they cannot contribute to the
$I=2$ amplitude and no CP violation is possible in the
$\pi^\pm\pi^0$ final states. This is identical to the situation
in $B^\pm \ra \rho^\pm \rho^0$. 
The penguin contribution
in $B^0 \ra \pi^+ \pi^-$, or in 
${\overline B}^0 \ra \pi^+ \pi^-$, 
can then be isolated and
removed by determining the relative magnitude and phase of the 
$I=0$ to $I=2$ amplitudes. We have 
\be
r_{\pi^+\pi^-}= e^{-2i\phi_m} {({\overline A}_2 - {\overline A}_0)
\over
({A}_2 - {A}_0)} = e^{2i\alpha} {(1 - {\overline z})\over (1 - z)}\;,
\label{rdef}
\ee
where $z ({\overline z}) \equiv A_0/A_2 ({\overline A}_0/{\overline A}_2)$
and ${\overline A}_2/A_2 = \exp(-2i \phi_t)$ with
$\phi_t \equiv {\rm arg} ( V_{ud} V_{ub}^\ast)$ and 
$\phi_m + \phi_t = \beta + \gamma = \pi - \alpha$ in the 
standard model~\cite{pdg96}.  
Given 
$|A^{+-}|$, $|A^{00}|$, 
$|A^{+0}|$, and their charge conjugates,
the measurement of ${\rm Im}\, r_{\pi^+\pi^-}$ determines $\sin 2\alpha$,
modulo discrete ambiguities in ${\rm arg} ((1-{\overline z})/(1-z))$.
The latter can be removed via a measurement of 
${\rm Im}\, r_{\pi^0\pi^0}$ as well~\cite{GL90}. 

We examine the manner in which
isospin-violating effects impact this extraction of 
$\sin 2\alpha$, for isospin 
is broken not only by 
electroweak effects but also by 
the $u$ and $d$ quark mass difference. 
Both sources of isospin violation generate $\Delta I=3/2$
penguin contributions, but the latter also generates 
$\pi^0-\eta,\eta'$ mixing~\cite{leut96}, admitting an $I=1$ 
amplitude. Although electroweak penguins 
are estimated 
to be small~\cite{deshe95,gronau95,fleischer96}, 
other isospin-violating 
effects, such as \pieta mixing, can also be important~\cite{pieta98,epsprime}.

To include the effects of \pieta mixing, we write the
pion mass eigenstate $|\pi^0\rangle$ 
in terms of the $SU(3)_f$ perfect states
$|\phi_3\rangle=|u{\overline u} - d{\overline d}\rangle/\sqrt{2}$, 
$|\phi_8\rangle=|u{\overline u} +
 d{\overline d} - 2s{\overline s}\rangle/\sqrt{6}$, 
and 
$|\phi_0\rangle=
|u{\overline u} + d{\overline d} + s{\overline s}\rangle/\sqrt{3}$.
To leading order in isospin violation~\cite{leut96} 
\be
|\pi^0 \rangle = |\phi_3\rangle + \varepsilon 
(\cos \theta |\phi_8\rangle - \sin \theta |\phi_0\rangle)
+ \varepsilon' (\sin \theta |\phi_8\rangle + \cos \theta
|\phi_0\rangle) \;,
\label{physpi}
\ee
where $|\eta\rangle=\cos \theta |\phi_8\rangle - \sin \theta |\phi_0\rangle
+ O(\varepsilon)$, and $|\eta'\rangle=\sin \theta |\phi_8\rangle + 
\cos \theta |\phi_0\rangle + O(\varepsilon')$. 
Expanding QCD to leading order in $1/N_c$, 
momenta, and quark masses and 
diagonalizing the
quadratic terms in $\phi_3$, $\phi_8$, and $\phi_0$ of the
resulting effective Lagrangian
determines 
the mass eigenstates $\pi^0$, $\eta$,
and $\eta'$ and yields
$\varepsilon = \varepsilon_0 \chi \cos \theta$ and 
$\varepsilon' = -2\varepsilon_0 \tilde\chi  \sin \theta$,
where $\chi= 1 + (4m_K^2 - 3m_{\eta}^2 - m_{\pi}^2)/(m_\eta^2 - m_\pi^2)
\approx 1.23$, $\tilde\chi = 1/\chi$, 
$\varepsilon_0 \equiv \sqrt{3}(m_d - m_u) /(4(m_s - \hat{m}))$,
 and $\hat{m}\equiv(m_u + m_d)/2$~\cite{leut96}. Thus the magnitude
of isospin breaking is controlled by the SU(3)-breaking parameter 
$m_s - \hat{m}$. The $\eta$-$\eta'$ mixing angle $\theta$ 
is found to be
$\sin 2\theta= - (4\sqrt{2}/3)(m_K^2 - m_\pi^2)/(m_{\eta'}^2 - m_\eta^2)
\approx -22^{\circ}$~\cite{leut96}. 
The resulting $\varepsilon
=1.14\varepsilon_0$ is comparable
to the one-loop-order chiral perturbation theory 
result of $\varepsilon=1.23\varepsilon_0$
in $\eta\ra \pi^+ \pi^- \pi^0$~\cite{gasser85,leut96}.
Using $m_q(\mu=2.5\, {\rm GeV})$ of Ali {\it et al.}~\cite{ali98}, we find
$\varepsilon = 1.4 \cdot 10^{-2}$ and
$\varepsilon' = 7.7 \cdot 10^{-3}$.

In the presence of isospin-violating effects, the \Bpipi amplitudes
become 
%
\bea
 A^{-0}&=& \langle \pi^- \phi_3 | {\cal H}^{\rm eff} | B^-\rangle 
+ \varepsilon_8\langle \pi^- \phi_{8} | 
{\cal H}^{\rm eff} | B^-\rangle 
+ \varepsilon_0\langle \pi^- \phi_{0} | 
{\cal H}^{\rm eff} | B^-\rangle \\
{\overline A}^{\,00}&=& \langle \phi_3 \phi_3 | {\cal H}^{\rm eff} | 
{\overline B}^0 \rangle 
+ 2\varepsilon_8\langle \phi_3 \phi_{8} | 
{\cal H}^{\rm eff} | {\overline B}^0 \rangle 
+ 2\varepsilon_0\langle \phi_3 \phi_{0} | 
{\cal H}^{\rm eff} | {\overline B}^0 \rangle \;,
\eea
\label{amps}
%
where $\varepsilon_8\equiv \varepsilon \cos\theta + \varepsilon' \sin\theta$
and 
$\varepsilon_0\equiv \varepsilon' \cos\theta - \varepsilon \sin\theta$. 
The $B\ra \pi\pi$ amplitudes satisfy
\bea
{\overline A}^{\,+-} + 2{\overline A}^{\,00} 
&-& \sqrt{2}\,A^{\,-0} 
= 
4\varepsilon_8 \langle \phi_3 \phi_{8} | 
{\cal H}^{\rm eff} | {\overline B}^0\rangle 
+4\varepsilon_0 \langle \phi_3 \phi_{0} | 
{\cal H}^{\rm eff} | {\overline B}^0\rangle \nonumber \\
&-& \sqrt{2} \varepsilon_8
\langle \pi^- \phi_{8} | 
{\cal H}^{\rm eff} | B^-\rangle 
- \sqrt{2} \varepsilon_0 
\langle \pi^- \phi_{0} | 
{\cal H}^{\rm eff} | B^-\rangle \;,
\label{newrel}
\eea
and thus the 
triangle relation implied by Eq.~\ref{triangle} becomes a quadrilateral.
We ignore 
the small mass
differences $m_{\pi^{\pm}}-m_{\pi^0}$ and $m_{B^{\pm}}-m_{B^0}$. 
\begin{figure}
\centerline{\epsfig{file=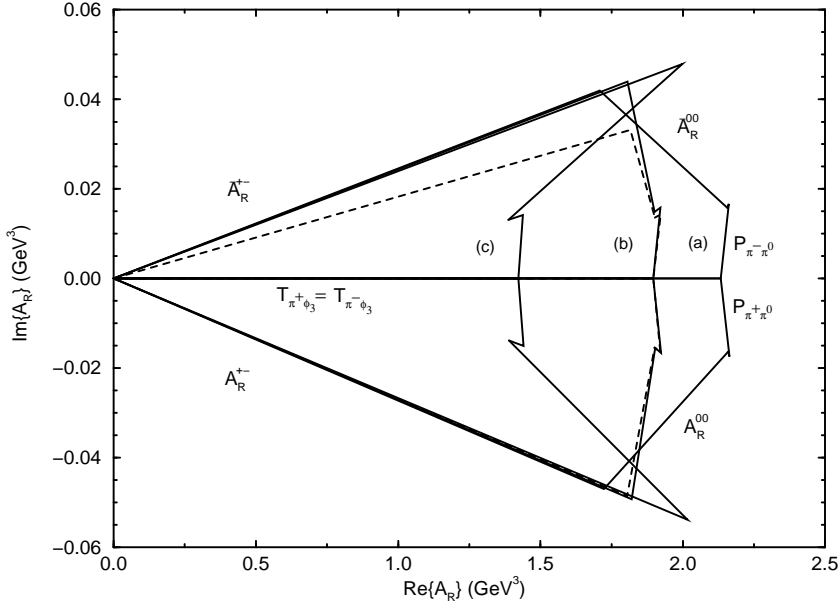,angle=-90,width=2.9in}}
\vspace{0.4in}
\caption{
Reduced amplitudes 
in $B\ra\pi\pi$ 
in the factorization 
approximation with 
[$N_c$, $k^2/m_b^2$] for  
a) [2,0.5], b) [3, 0.5] (solid line) and [3, 0.3] (dashed line), and
c) [$\infty$,0.5]. 
}
 \label{figtri}
\vspace{-5mm}
\end{figure}

  We proceed by computing the individual amplitudes using the 
$\Delta B=1$ effective Hamiltonian resulting from the operator
product expansion in 
QCD in NLL order~\cite{ali98}, using 
the factorization approximation for the hadronic matrix elements.
In this context, 
we can then apply the isospin analysis delineated above to
infer $\sin 2\alpha$ and thus estimate its theoretical systematic error, 
incurred through the neglect of isospin violating
effects. 
Numerical results 
for the reduced amplitudes $A_R$ and ${\overline A}_R$, where
${\overline A}_R^{\,00} \equiv 
2{\overline A}^{\,00} /((G_F/\sqrt{2}) i V_{ub} V_{ud}^*)$,
${\overline A}_R^{\,+-} \equiv 
{\overline A}^{\,+-} /((G_F/\sqrt{2}) i V_{ub} V_{ud}^\ast)$, and
$A_R^{-0} \equiv \sqrt{2}A^{-0} /((G_F/\sqrt{2}) i V_{ub} V_{ud}^\ast)$,
with $N_c=2,3,\infty$ and $k^2/m_b^2=0.3,0.5$ 
are shown in Fig.~2. 
$A_R^{+0}$ and $A_R^{-0}$ are
broken into tree and penguin contributions, so that 
$A_R^{+0}\equiv T_{\pi^+\phi_3} + P_{\pi^+\pi^0}$ and 
$A_R^{-0}\equiv T_{\pi^-\phi_3} + P_{\pi^-\pi^0}$, where 
$P_{\pi^\pm\pi^0}$ is defined to include the isospin-violating tree
contribution in $A_R^{\pm0}$ as well. The shortest side in each
polygon is the vector defined by the RHS of Eq.~\ref{newrel}.
The values of $\sin2\alpha$ extracted from the computed amplitudes 
with $N_c$ and $k^2/m_b^2=0.5$ are shown
in Table \ref{delcf} --- the results for $k^2/m_b^2=0.3$ are similar
and have been omitted.
%
\setcounter{footnote}{0}
\begin{table}
    \caption{
Strong phases and inferred values of $\sin2\alpha$~\protect{\cite{GL90}} 
from amplitudes in the factorization
approximation with $N_c$ and $k^2/m_b^2=0.5$. 
The strong phase $2\delta_{\rm true}$
is the opening
angle between the ${\overline A_R}^{+-}$ and ${A_R}^{+-}$ amplitudes
in Fig.~1, 
whereas 2$\delta_{\rm GL}$ is the strong phase associated
with the closest matching $\sin2\alpha$ values, denoted 
$(\sin 2\alpha)_{\rm GL}$, from 
Im$r_{\pi^+\pi^-}$/Im$r_{\pi^0\pi^0}$, respectively. 
The bounds $|2\delta_{\rm GQI}|$ and $|2\delta_{\rm GQII}|$  on 
$2\delta_{\rm true}$ 
from Eqs.~2.12 and 2.15 of Ref.~\protect{\cite{GQpi97}} are also shown.
All angles are in degrees.
We input a) $\sin2\alpha=0.0432$~\protect{\cite{ali98,SMparam,mele98}},
b) $\sin2\alpha=-0.233$ ($\rho=0.2,\eta=0.35$)~\protect{\cite{DLcomm}}, and 
c) $\sin2\alpha=0.959$ ($\rho=-0.12$).}
\label{delcf}
\vspace{3mm}
 \begin{center}
\begin{tabular}{|ccccccc|}
\hline
case & $N_c$ & 2$\delta_{\rm true}$ &  $|2\delta_{\rm GQI}|$ & 
$|2\delta_{\rm GQII}|$ &  $|2\delta_{\rm GL}|$ &  $(\sin 2\alpha)_{\rm GL}$  \\
\hline
a & 2 &  24.4 & 26.1 & 15.8 & 16.6 & -0.0900/-0.0221 \\
a & 3 &  24.2 & 16.9 & 16.1  & 16.2  & -0.0926/0.107 \\
a & $\infty$ & 23.8  & 59.4  & 25.1  & 23.6 & 0.0451/0.394 \\
b & 2 &  19.6 & 23.4 & 12.1 & 12.9 & -0.343/-0.251 \\
b & 3 &  19.4 & 13.5 & 12.9  & 13.0  & -0.719/-0.855($\ast$)\\
b & $\infty$ & 19.2  & 59.9  & 23.6  & 0.76 & -0.550/-0.814($\ast,\dagger$) \\
c & 2 &  28.3 & 36.5 & 20.4 & 21.0 & 0.917/0.915 \\
c & 3 &  28.0 & 24.0 & 19.1  & 19.0  & 0.905/0.952 \\
c & $\infty$ & 28.3  & 36.5  & 20.4  & 21.0 & 0.917/0.915 \\
\hline
\end{tabular}
\end{center}
\setcounter{footnote}{0}
\vspace{3mm}
\baselineskip=8 pt
{\footnotesize
$^\ast$ The matching
procedure fails to choose a $\sin2\alpha$ which is as 
close to the input value as possible. \\
$^\dagger$ The discrete ambiguity in the strong phase is resolved wrongly.}
\vspace{-7mm}
\end{table}
In the presence of $\pi^0$-$\eta,\eta'$ mixing,
the ${\overline A}_R^{\,+-}$, ${\overline A}_R^{\,-0}$, and 
${\overline A}_R^{\,00}$ amplitudes obey a quadrilateral relation 
as per Eq.~\ref{newrel}.
Consequently, 
the values of $\sin2\alpha$ 
extracted from Im$r_{\pi^+\pi^-}$ and Im$r_{\pi^0\pi^0}$ can not only differ
markedly from the value of $\sin2\alpha$ input but also need not match.
The incurred error in $\sin 2\alpha$ increases as the value 
to be extracted decreases; the structure of Eq.~\ref{rdef} suggests this,
for as $\sin 2\alpha$ decreases the quantity
Im$((1 - {\overline z})/(1 -z))$ becomes more important to 
determining the extracted value. 
It is useful to constrast the impact of 
the various isospin-violating effects.
The presence of $\Delta I=3/2$ penguin
contributions, be they from $m_u\ne m_d$ or electroweak effects, shift
the extracted value of $\sin 2\alpha$ from its input value, yet the 
``matching'' of the $\sin 2\alpha$ values in
Im$r_{\pi^+\pi^-}$ 
and Im$r_{\pi^0\pi^0}$ is
unaffected. The mismatch troubles seen in Table \ref{delcf} are driven by
$\pi^0$-$\eta,\eta'$ mixing, though the latter shifts the values of
$\sin 2\alpha$ in Im$r_{\pi^+\pi^-}$ as well. 
Picking the closest
matching values of $\sin 2\alpha$ in the two final states also 
picks the solutions closest to the input value; the exceptions are noted
in Table \ref{delcf}. 
If $|A^{00}|$ and $|{\overline A}^{00}|$ 
are small~\cite{GL90} the complete isospin analysis may not be
possible, so that 
we also examine the utility of the bounds recently proposed by Grossman and
Quinn~\cite{GQpi97}
on the strong phase 
$2\delta_{\rm true}\equiv {\rm arg}((1 - {\overline z})/(1 - z))$
of Eq.~\ref{rdef}. The bounds 
$2\delta_{\rm GQI}$ and $2\delta_{\rm GQII}$ given by their~{\cite{GQpi97}} 
Eqs.~2.12 and 2.15, respectively, 
follow from Eq.~\ref{triangle}, and thus can be broken
by isospin-violating effects. 
As shown in 
Table \ref{delcf}, the bounds typically are broken, and their 
efficacy does not improve as the value of $\sin 2\alpha$ 
to be reconstructed grows large. 
  
  To conclude, we have considered the role of isospin violation
in $B\ra\pi\pi$ decays and have found the effects to be significant. 
Most particularly, the utility of the
isospin analysis in determining $\sin 2\alpha$ strongly depends 
on the value to be reconstructed. The error in $\sin 2\alpha$
from a Im$r_{\pi^+\pi^-}$ measurement can be 50\% or more for
the small values of
$\sin 2 \alpha$ currently favored by 
phenomenology~\cite{ali98,SMparam,mele98,DLcomm}; however, 
if $\sin 2\alpha$ were
near unity, the error would decrease to less than 10\%. 
The effects found arise in part because the penguin contribution in 
$B^0 \rightarrow \pi^+ \pi^-$, e.g., is itself small; we estimate
$|P|/|T| < 9\% | V_{tb}V^*_{td}| / |V_{ub}V^*_{ud}|$. Relative to this
scale, the impact of \pieta mixing is significant. Yet, 
were the penguin
contributions in $B\rightarrow \pi\pi$ larger, the isospin-violating
effects considered would still be germane, for not only would 
the $\Delta I=3/2$ penguin contributions likely be larger but 
the $B\rightarrow \pi\eta$ and
$B\rightarrow \pi\eta'$ contributions could also be 
larger as well~\cite{ciuchini97}. 
To conclude, we have shown that 
the presence of $\pi^0$-$\eta,\eta'$ mixing breaks the
triangle relationship, Eq.~\ref{triangle}, usually assumed~\cite{GL90} 
and can mask the true value of $\sin2\alpha$.

%
%
\vspace{-3mm}
\section*{Acknowledgements}

I am grateful to 
H.~B. O'Connell and A.~W. Thomas 
for their collaboration on issues
related to \rhoom mixing and to 
the organizers for their invitation to speak at this workshop. 
This work was supported by the U.S. Department of Energy 
under DE-FG02-96ER40989 and by the Special Research Centre for
the Subatomic Structure of Matter at the University of Adelaide. 

%
%

\end{document}